\documentclass[11pt]{llncs}

\usepackage{fullpage}

\usepackage{amssymb}

\usepackage{algorithm2e}

\newcommand{\size}[1]{{\ensuremath{\mathrm{size}\left(#1\right)}}}
\newcommand{\Gal}[1]{{\ensuremath{\mathrm{Gal}\left(#1\right)}}}
\newcommand{\Blocks}[1]{{\ensuremath{\mathcal{B}\left(#1\right)}}}
\newcommand{\Gof}[1]{{\ensuremath{\mathrm{G}\left(#1\right)}}}

\newcommand{\Sym}[1]{{\ensuremath{\mathrm{Sym}\left(#1\right)}}}
\newcommand{\pr}[2]{{\ensuremath{\left.{#1}\right\vert_{#2}}}}

\newcommand{\Q}{\ensuremath{\mathbb{Q}}}

\title{A Polynomial Time Nilpotence Test for Galois Groups and 
Related Results}

\author{V.~Arvind\inst{1} \and
  Piyush P Kurur\inst{2}
}
\institute{
  Institute of Mathematical Sciences\\ % 
  C.I.T Campus,Chennai, India 600 113\\% 
  \email{arvind@imsc.res.in}% 
  \and
  Department of Computer Science and Engineering,\\
  Indian Institute of Technology, Kanpur,\\
  Kanpur, UP 208016, India\\
  \email{ppk@cse.iitk.ac.in}
\thanks{work done when the author was a PhD
    student at the Institute of Mathematical Sciences, Chennai.}
}
\date{} 

\begin{document}
\maketitle
\begin{abstract}
  We give a deterministic polynomial-time algorithm to check whether
  the Galois group $\Gal{f}$ of an input polynomial $f(X) \in \Q[X]$
  is nilpotent: the running time is polynomial in $\size{f}$. Also, we
  generalize the Landau-Miller solvability test to an algorithm that
  tests if $\Gal{f}$ is in $\Gamma_d$: this algorithm runs in time
  polynomial in $\size{f}$ and $n^d$ and, moreover, if
  $\Gal{f}\in\Gamma_d$ it computes all the prime factors of $\#
  \Gal{f}$.
\end{abstract}

\section{Introduction}
Computing the Galois group of a polynomial is a fundamental problem in
algorithmic number theory. Asymptotically, the best known algorithm is
due to Landau~\cite{landau84galois}: on input $f(X)$, it takes time
polynomial in $\size{f}$ and the order of its Galois group $\Gal{f}$.
If $f(X)$ has degree $n$ then $\Gal{f}$ can have $n!$ elements. Thus,
Landau's algorithm takes time exponential in input size. It is a long
standing open problem if there is an asymptotically faster algorithm
for computing $\Gal{f}$. Lenstra's survey \cite{lenstra92algorithm}
discusses this and related problems.

A different kind of problem is to test for a given $f(x)$ if $\Gal{f}$
satisfies a specific property without explicitly computing it.
Galois's seminal work showing $f(X)$ is solvable by radicals if and
only if $\Gal{f}$ is solvable is a classic example. Landau and Miller
\cite{landau85solvability} gave a remarkable polynomial-time algorithm
for testing solvability of the Galois group without computing the
Galois group.

\subsection{\bf The results of this article}

Our main result is a deterministic polynomial-time algorithm for
testing if $\Gal{f}$ is nilpotent. Although nilpotent groups are a
proper subclass of solvable groups, the Landau-Miller solvability test
does not give a nilpotence test. Basically, the Landau-Miller test is
a method of testing that all composition factors of $\Gal{f}$ are
abelian, which tests solvability. Nilpotence however is a more
``global'' property, in the sense that it cannot be inferred by
properties of the composition factors alone.

We note here that nilpotence testing of Galois groups has been
addressed by other researchers with the goal of developing good
practical algorithms. For example in ~\cite{FG03} an algorithm for
nilpotence testing is given which takes time polynomial in $\size{f}$
and $\# \Gal{f}$. However, ours is the first algorithm that is
provably polynomial time, i.e. runs in time polynomial in $\size{f}$,
on all inputs.

Next, we show that the Landau-Miller solvability test can be extended
to a polynomial-time algorithm for checking, given $f\in\Q[X]$, if
$\Gal{f}$ is in $\Gamma_d$ for constant $d$. A group $G$ is in
$\Gamma_d$ if there is a composition series $G = G_0 \rhd \ldots \rhd
G_t = \{ 1 \}$ such that each nonabelian composition factor
$G_i/G_{i+1}$ is isomorphic to a subgroup of $S_d$. The class
$\Gamma_d$ often arises in permutation group algorithms (see
e.g.~\cite{luks82bounded}). Moreover, if $\Gal{f}\in\Gamma_d$, the
prime factors of $\# \Gal{f}$ can be found in polynomial time.

\subsection{\it Galois theory overview}

We quickly recall some Galois theory (see, e.g.~\cite{lang:algebra}
for details).  Let $L$ and $K$ be fields. If $L\supset K$, we say that
$L$ is an extension of $K$ and denote it by $L/K$. If $L/K$ then $L$
is a vector space over $K$ and by the \emph{degree} of $L/K$, denoted
by $[L:K]$, we mean its dimension. An extension $L/K$ is \emph{finite}
if its degree $[L:K]$ is finite. If $L/M$ and $M/K$ are finite
extensions then $[L:K] = [L:M].[M:K]$. The polynomial ring $K[X]$ is a
unique factorisation domain: every polynomial can be uniquely (upto
scalars) written as a product of irreducible polynomials. Let $L/K$ be
an extension. An $\alpha \in L$ is \emph{algebraic} over $K$ if
$f(\alpha)=0$ for some $f(X)\in K[X]$. For $\alpha$ algebraic over
$K$, the \emph{minimal polynomial} of $\alpha$ over $K$ is the unique
monic polynomial $\mu_\alpha[K](X)$ of least degree in $K[X]$ for
which $\alpha$ is a root. We write $\mu_\alpha(X)$ for
$\mu_\alpha[K](X)$ when $K$ is understood. Elements $\alpha,\beta\in
L$ are \emph{conjugates} over $K$ if they have the same minimal
polynomial over $K$. The smallest subfield of $L$ containing $K$ and
$\alpha$ is denoted by $K(\alpha)$.

The \emph{splitting field} $K_f$ of $f\in K[X]$ is the smallest
extension of $K$ containing all the roots of $f$. A finite extension
$L/K$ is \emph{normal} if for all irreducible polynomials $f(X)\in
K[X]$, either $f(X)$ splits or has no root in $L$. Any normal
extension over $K$ is the splitting field of some polynomial in
$K[X]$. An extension $L/K$ is \emph{separable} if for all irreducible
polynomials $f(X) \in K[X]$ there are no multiple roots in $L$.  A
normal and separable finite extension $L/K$ is a \emph{Galois
  extension}.

The \emph{Galois group} $\Gal{L/K}$ of $L/K$ is the subgroup of
automorphisms $\sigma$ of $L$ that leaves $K$ fixed, i.e.\
$\sigma(\alpha) = \alpha$ for all $\alpha \in K$. The Galois group
$\Gal{f}$ of $f\in K[X]$ is $\Gal{K_f/K}$. For a subgroup $G$ of
automorphisms of $L$, the \emph{fixed field} $L^G$ is the largest
subfield of $L$ fixed by $G$.  We now state the fundamental theorem of
Galois.

\begin{theorem}{\rm{\cite[Theorem 1.1, Chapter
      VI]{lang:algebra}}}\label{thm-funda-galois} Let $L/K$ be a
  Galois extension with Galois group $G$. There is a one-to-one
  correspondence between subfields $E$ of $L$ containing $K$ and
  subgroups $H$ of $G$, given by $E \rightleftharpoons L^H$.  The
  Galois group of $\Gal{L/E}$ is $H$ and $E/K$ is a Galois extension
  if and only if $H$ is a normal subgroup of $G$. If $H$ is a normal
  subgroup of $G$ and $E = L^H$ then $\Gal{E/K}$ is isomorphic to the
  quotient group $G/H$.
\end{theorem}

\subsection{\it Presenting algebraic numbers, number fields and Galois
  groups}

The algorithms we describe take objects like algebraic numbers, number
fields etc. as input. We define sizes of these objects. Integers are
encoded in binary. A rational $r$ is given by coprime integers $a,b$
such that $r =a/b$. Thus, $\size{r}$ is $\size{a}+ \size{b}$.  A
polynomial $T(X) = a_0 + \ldots + a_n X^n \in\Q[X]$ is given by a list
of its coefficients. Thus, $\size{T}$ is defined as $\sum \size{a_i}$.

%% edited by ppk 30 march 2006
%% moved the primitive element theorem here as we do not need the
%% result in full generality and see the comment of the referee.
%% In particular for non-separable fields primitive element theorem
%% could be violated.
%%

A \emph{number field} is a finite extension of $\Q$. Let $K/\Q$ be a
number field of degree $n$.  By the primitive element theorem
\cite[Theorem 4.6, Chapter V]{lang:algebra}, there is an algebraic
number $\eta\in K$ such that $K = \Q(\eta)$. Such an element is a
\emph{primitive element} of $K/\Q$ and its minimal polynomial is a
\emph{primitive polynomial}. Let $\mu_\eta(X)$ be the minimal
polynomial of $\eta$ over $\Q$. Then the field $K$ can be written as
the quotient $K=\Q[X]/\mu_\eta(X)$. Thus $K$ can be presented by
giving a primitive polynomial for $K/\Q$. We can assume that $\eta$ is
an algebraic integer and hence its minimal polynomial $\mu_\eta(X)$
has integer coefficients~\cite[Proposition 1.1, Chapter
VII]{lang:algebra}.  When we say that an algorithm takes a number
field $K$ as input we mean that it takes a primitive polynomial
$\mu_\eta(X)$ for $K$ as input. Thus the input size for $K$, which we
denote by $\size{K}$, is defined to be $\size{\mu_\eta}$.

Suppose $K=\Q(\eta)$ is presented by $\mu_\eta(X)$.  Notice that each
$\alpha\in K$ can be expressed as $\alpha = A_\alpha(\eta)$ for a
unique polynomial $A_\alpha(X) \in \Q[X]$ of degree less than $n$. By
$\size{\alpha}$ we mean $\size{A_\alpha(X)}$.  Note that the size of
$\alpha \in K$ depends on the primitive element $\eta\in K$. Now, for
a polynomial $f(X) = a_0 + \ldots + a_m X^m$ in $K[X]$ we define
$\size{f}$ to be $\sum \size{a_i}$.

Let $f(X) \in \Q[X]$ of degree $n$. For an algorithm purporting to
compute $\Gal{f}$, one possibility is that it outputs the complete
multiplication table for $\Gal{f}$. However, this could be exponential
in $\size{f}$ as $\Gal{f}$ can be as large as $n!$. A succinct
presentation of $\Gal{f}$ is as a permutation group acting on the
roots of $f$ since elements of $\Gal{f}$ permute the roots of $f$ and
are completely determined by their action on the roots of $f$. Thus,
by numbering the roots of $f$, we can consider $\Gal{f}$ as a subgroup
of the symmetric group $S_n$ (note here that $\Gal{f}$ is determined
only up to conjugacy as the numbering of the roots is arbitrary).
Since any subgroup of $S_n$ has a generator set of size $n-1$ (see
e.g.\ \cite{luks93permutation}), we can present $\Gal{f}$ in size
polynomial in $n$. Thus, by computing $\Gal{f}$ we mean finding a
small generator set for it as a subgroup of $S_n$. Determining
$\Gal{f}$ as a subgroup of $S_n$ is a reasonable way of describing the
output. Algorithmically, we can answer several natural questions about
a subgroup $G$ of $S_n$ given by generator set in polynomial time.
E.g.\ testing if $G$ is solvable, finding a composition series for $G$
etc.\ \cite{luks93permutation}.

\subsubsection*{\it Previous complexity results}

As mentioned, the best known algorithm for computing the Galois group
of a polynomial is due to Landau~\cite{landau84galois}.

\begin{theorem}[Landau] \label{thm-landau-galois-algo} There is a
  deterministic algorithm that takes as input a number field $K$, a
  polynomial $f(X) \in K[X]$ and a positive integer $b$ in unary, and
  in time bounded by $\size{f}$, $\size{K}$ and $b$, decides if
  $\Gal{K_f/K}$ has at most $b$ elements, and if so computes
  $\Gal{K_f/K}$ by finding the entire multiplication table of
  $\Gal{K_f/K}$ (and hence also by giving the generating set of
  $\Gal{K_f/K}$ as a permutation group on the roots of $f(X)$).
\end{theorem}

The algorithm first computes a primitive element $\theta$ of $K_f$.
Determining $\Gal{f}$ amounts to finding the action of the
automorphisms on $\theta$. Subsequently, Landau and Miller
\cite{landau85solvability} gave their polynomial-time solvability
test.

\begin{theorem}[Landau-Miller]\label{thm-landau-solvability}
  Given $f(X) \in \Q[X]$ there is a deterministic polynomial-time
  algorithm for testing if $\Gal{f}$ is solvable.
\end{theorem}

\section{Preliminaries}
We recall some permutation group theory from Wielandt's
book~\cite{wielandt64finite}. Let $\Omega$ be a finite set. The
\emph{symmetric group} $\Sym{\Omega}$ is the group of all permutations
on $\Omega$.  By a \emph{permutation group on $\Omega$} we mean a
subgroup of $\Sym{\Omega}$. For $\alpha \in \Omega$ and $g \in
\Sym{\Omega}$, let $\alpha^g$ denote the image of $\alpha$ under the
permutation $g$. For $A \subseteq \Sym{\Omega}$, $\alpha^A$ denotes
the set $\{ \alpha^g : g \in A\}$. In particular, for
$G\leq\Sym{\Omega}$ the \emph{$G$-orbit} containing $\alpha$ is
$\alpha^G$. The $G$-orbits form a partition of $\Omega$.  Given
$G\leq\Sym{\Omega}$ by a generating set $A$ and $\alpha \in \Omega$,
there is a polynomial-time algorithm to compute
$\alpha^G$~\cite{luks93permutation}.

For $\Delta \subseteq \Omega$ and $g \in \Sym{\Omega}$, $\Delta^g$
denotes $\{ \alpha^g : \alpha \in \Delta \}$. The setwise stabilizer
of $\Delta$, i.e. $\{ g \in G : \Delta^g = \Delta\}$, is denoted by
$G_\Delta$. If $\Delta$ is the singleton set $\{ \alpha \}$ we write
$G_\alpha$ instead of $G_{\{\alpha\}}$.  For any $\Delta$ by
$\pr{G}{\Delta}$ we mean $G_\Delta$ restricted to $\Delta$. An often
used result is the orbit-stabilizer formula stated below~\cite[Theorem
3.2]{wielandt64finite}.

\begin{theorem}[Orbit-stabilizer formula]
  Let $G$ be a permutation group on $\Sym{\Omega}$ and let $\alpha$ be
  any element of $\Omega$ then the order of the group $G$ is given by
  $\# G = \# G_\alpha . \# \alpha^G$.
\end{theorem}

A permutation group $G$ on $\Omega$ is \emph{transitive} if there is a
single $G$-orbit. Suppose $G \leq \Sym{\Omega}$ is transitive.  Then
$\Delta\subseteq \Omega$ is a \emph{$G$-block} if for all $g \in G$
either $\Delta^g = \Delta$ or $\Delta^g \cap \Delta=\emptyset$.  For
every $G$, $\Omega$ is a block and each singleton $\{\alpha\}$ is a
block. These are the \emph{trivial blocks} of $G$. A transitive group
$G$ is \emph{primitive} if it has only trivial blocks and it is
\emph{imprimitive} if it has nontrivial blocks. A $G$-block $\Delta$
is a \emph{maximal subblock} of a $G$-block $\Sigma$ if $\Delta
\subset \Sigma$ and there is no $G$-block $\Upsilon$ such that $\Delta
\subset \Upsilon \subset \Omega$. Let $\Delta$ and $\Sigma$ be two
$G$-blocks. A chain $\Delta = \Delta_0 \subset \ldots \subset \Delta_t
= \Sigma$ is a \emph{maximal chain} of $G$-blocks between $\Delta$ and
$\Sigma$ if for all $i$, $\Delta_i$ is a maximal subblock of
$\Delta_{i+1}$.

For a $G$-block $\Delta$ and $g \in G$, $\Delta^g$ is also a $G$-block
such that $\# \Delta = \# \Delta^g$. Let $\Delta$ and $\Sigma$ be two
$G$-blocks such that $\Delta \subseteq \Sigma$.  The
\emph{$\Delta$-block system of $\Sigma$}, is the collection
\[
\Blocks{\Sigma/\Delta} = \{ \Delta^g : g \in G \textrm{ and } \Delta^g
\subseteq \Sigma \}.
\] 
The set $\Blocks{\Sigma/\Delta}$ is a partition of $\Sigma$.  It
follows that $\# \Delta$ divides $\# \Sigma$ and by \emph{index} of
$\Delta$ in $\Sigma$, which we denote by $[\Sigma:\Delta]$, we mean
$\# \Blocks{\Sigma/\Delta} = \frac{\# \Sigma}{\# \Delta}$. We will use
$\Blocks{\Delta}$ to denote $\Blocks{\Omega/\Delta}$. We state the
connection between blocks and subgroups \cite[Theorem
7.5]{wielandt64finite}.

\begin{theorem}[Galois correspondence of blocks]\label{thm-blocks-galois}
  Let $G\leq\Sym{\Omega}$ be transitive and $\alpha\in\Omega$. For $G
  \geq H \geq G_\alpha$ the orbit $\Delta = \alpha^H$ is a $G$-block
  and $G_\Delta = H$. The correspondence $\alpha^H = \Delta
  \rightleftharpoons G_\Delta = H$ is a one-to-one correspondence
  between $G$-blocks $\Delta$ containing $\alpha$ and subgroups $H$ of
  $G$ containing $G_\alpha$. Furthermore for $G$-blocks $\Delta
  \subseteq \Sigma$ we have $[G_\Sigma : G_\Delta] = [\Sigma :
  \Delta]$.
\end{theorem}

Let $G\leq\Sym{\Omega}$ be transitive and $\Delta$ and $\Sigma$ be two
$G$-blocks such that $\Delta\subseteq \Sigma$. Let $G(\Sigma/\Delta)$
denote the group $\{ g \in G : \Upsilon^g = \Upsilon \textrm{ for all
} \Upsilon \in \Blocks{\Sigma/\Delta} \}$.  We write $G^\Delta$ for
the group $\Gof{\Omega/\Delta}$.  For any $g \in G_\Sigma$, since $g$
setwise stabilises $\Sigma$, $g$ permutes the elements of
$\Blocks{\Sigma/\Delta}$. Hence for any $\Upsilon \in
\Blocks{\Sigma/\Delta}$ we have $\Upsilon^{g^{-1}\Gof{\Sigma/\Delta}g}
= \Upsilon$. Thus, $\Gof{\Sigma/\Delta}$ is a normal subgroup of
$G_\Sigma$.  In particular, $G^\Delta$ is a normal subgroup of $G$.

\noindent{\bf Remark.}~~The following two lemmata are quite standard
in permutation group theory. For the reader's convenience we have
included short proofs. The following lemma lists important properties
of $G^\Delta$.

\begin{lemma}\label{lem-gsupdelta}\hfill{~}
  \begin{enumerate}
  \item For a $G$-block $\Delta \subseteq \Sigma$,
    $\Gof{\Sigma/\Delta}$ is the largest normal subgroup of $G_\Sigma$
    contained in $G_\Delta$.

  \item Let $\Sigma$ be $G$-block then $G^{\Sigma} \hookrightarrow
    \prod_{\Upsilon \in \Blocks{\Sigma}} \pr{G}{\Upsilon}$.

  \item Let $\Delta$ be a $G$-subblock of $\Sigma$ then
    $\frac{G_\Sigma}{\Gof{\Sigma/\Delta}}$ is a faithful permutation
    group on $\Blocks{\Sigma/\Delta}$ and is primitive when $\Delta$
    is a maximal subblock.

  \item The quotient group $G^\Sigma/G^\Delta$ can be embedded as a
    subgroup of $\left(\frac{G_\Sigma}{\Gof{\Sigma/\Delta}}\right)^l$
    for some $l$.

% More precisely the following is a group embedding:
%    \[
%    G^{\Sigma}/G^\Delta \hookrightarrow \prod_{\Upsilon \in
%      \Blocks{\Sigma}}
%    \frac{G_\Upsilon}{\Gof{\Upsilon/\Delta_{\Upsilon}}}.
%    \]
\end{enumerate}
\end{lemma}
\begin{proof}
  Let $N\subseteq G_\Delta$ be a normal subgroup of $G_\Sigma$. Since
  $\Delta^N = \Delta$, and since $G_\Sigma$ acts transitively on
  $\Blocks{\Sigma/\Delta}$, for any $\Upsilon \in
  \Blocks{\Sigma/\Delta}$ there is a $g \in G_\Sigma$ such that
  $\Upsilon = \Delta^g$. Therefore, $\Upsilon^N = \Delta^{gN} =
  \Delta^{Ng}= \Upsilon$ for each $\Upsilon \in
  \Blocks{\Sigma/\Delta}$. Thus $N\subseteq \Gof{\Sigma/\Delta}$.
  Since $\Gof{\Sigma/\Delta}\unrhd G_\Sigma$ we have proved part 1.

  Part 2 directly follows from the definition of $G^\Sigma$. Part 3
  follows from the fact that $g, h\in G_\Sigma$ have the same action
  on $\Blocks{\Sigma/\Delta}$ precisely when
  $g\Gof{\Sigma/\Delta}=h\Gof{\Sigma/\Delta}$. The nontrivial
  $\frac{G_\Sigma}{\Gof{\Sigma/\Delta}}$-blocks of
  $\Blocks{\Sigma/\Delta}$ are in 1-1 correspondence with the
  $G$-blocks properly between $\Delta$ and $\Sigma$. Thus,
  $\frac{G_\Sigma}{\Gof{\Sigma/\Delta}}$ is primitive if and only if
  $\Delta$ is a maximal subblock of $\Sigma$.

  For Part 4 notice that we have the group isomorphism
  \[
  \frac{\pr{G}{\Upsilon}}{\pr{\Gof{\Upsilon/\Delta_{\Upsilon}}}{\Upsilon}}
  \cong \frac{G_\Upsilon}{\Gof{\Upsilon/\Delta_{\Upsilon}}},
  \]
  for each $\Upsilon\in\Blocks{\Sigma}$. As $G^\Delta = G^\Sigma \cap
  \prod \pr{\Gof{\Upsilon/\Delta_\Upsilon}}{\Upsilon}$ we have
  \[
  G^\Sigma /G^\Delta \hookrightarrow \prod_{\Upsilon \in
    \Blocks{\Sigma}}
  \frac{\pr{G}{\Upsilon}}{\pr{\Gof{\Upsilon/\Delta_{\Upsilon}}}{\Upsilon}}
  = \prod_{\Upsilon \in \Blocks{\Sigma}}
  \frac{G_\Upsilon}{\Gof{\Upsilon/\Delta_{\Upsilon}}}.\]

  Let $g \in G$ such that $\Delta^g=\Delta_{\Upsilon}$. Then,
  $G_\Upsilon = g^{-1} G_\Sigma g$ and $\Gof{\Upsilon/\Delta_\Upsilon}
  = g^{-1}\Gof{\Sigma/\Delta}g$. Thus,
  $\frac{G_\Sigma}{\Gof{\Sigma/\Delta}}$ and
  $\frac{G_\Upsilon}{\Gof{\Upsilon/\Delta_\Upsilon}}$ are isomorphic,
  which implies that $G^\Sigma/G^\Delta$ is isomorphic to a subgroup
  of $\left(\frac{G_\Sigma}{\Gof{\Sigma/\Delta}}\right)^l$ for some
  $l$.
\end{proof}

\begin{lemma}\label{lem-orbit-normal}
  Let $G\leq\Sym{\Omega}$ be transitive and $N\unlhd G$. Let
  $\alpha\in\Omega$. Then the $N$-orbit $\alpha^N$ is a $G$-block and
  the collection of $N$-orbits is an $\alpha^N$-block system of
  $\Omega$ under $G$ action. If $N\neq\{ 1 \}$ then $\|\alpha^N\|>1$.
  Furthermore, if $G_\alpha\leq N\neq G$ then the $\alpha^N$-block
  system is nontrivial implying that $G$ is not primitive.
\end{lemma}
\begin{proof}
  Let $\alpha\in\Omega$ and $g \in G$. Then $(\alpha^N)^g =
  \alpha^{Ng} = \alpha^{gN} = (\alpha^g)^N$. Thus $(\alpha^N)^g$ and
  $\alpha^N$ are $N$-orbits, and hence are identical or disjoint.
  Thus, $\alpha^N$ is a $G$-block and the $N$-orbits form a block
  system. Clearly, if $\alpha^N = \{ \alpha \}$ then $N=\{1\}$.
  Finally, by the Orbit-Stabilizer formula $\# G=\#\Omega\cdot\#
  G_\alpha$ and $\# N=\#\alpha^N\cdot\# G_\alpha$.  Thus, if
  $\{1\}\neq N\neq G$ then $\alpha^N$ is a proper $G$-block.
\end{proof}

\section{Nilpotence testing for Galois groups}

First we recall crucial properties of nilpotent transitive permutation
groups. These are standard group theoretic facts that we assemble
together and, for the sake of completeness, provide proof sketches
where necessary. We start with a characterization of finite nilpotent
groups. Let $G$ be a finite group and $p_1,\ldots,p_k$ be the prime
factors of $\#G$. For each $i$, let $G_{p_i}$ be a $p_i$-Sylow
subgroup of $G$. Then $G$ is \emph{nilpotent} if and only if $G$ is
the (internal) direct product $G_{p_1}\times\ldots \times G_{p_k}$.
Consequently, $G_{p_i}$ is the unique $p_i$-Sylow subgroup of $G$ for
each $i$ and hence $G_{p_i}\lhd G$.

\begin{lemma}\label{lem-struct-nilpotent}
  Let $G\leq\Sym{\Omega}$ be transitive and nilpotent, and $p$ be any
  prime. Then
  \begin{enumerate}
  \item[(1)] The prime $p$ divides $\#G$ if and only if $p$ divides
    $\# \Omega$.\label{lem-struct-nilpotent-ps}
  \item[(2)] If $p \mid \# G$ and $\alpha \in \Omega$ then there is a
    block $\Sigma_p^\alpha$ containing $\alpha$ such that $\#
    \Sigma_p^\alpha$ is the highest power of $p$ that divides $\#
    \Omega$.\label{lem-struct-nilpotent-sigma}
  \item[(3)] Let $\Delta$ be any $G$-block containing $\alpha$ such
    that $\# \Delta = p^l$ and suppose $p$ divides $\# G$. Then
    $\Delta \subseteq \Sigma_p^\alpha$. Also, for $q\neq p$, the
    $q$-Sylow subgroup of $G_\Delta$ is given by $G_q \cap G_\Delta =
    G_q \cap G_\alpha$.
  \end{enumerate}
\end{lemma}
\begin{proof}
  Part (1): As $G$ is transitive, $\# \Omega$ divides $\# G$. Hence,
  each prime factor of $\# \Omega$ divides $\# G$. Let $p$ be a prime
  factor of $\# G$. For $\alpha \in \Omega$, let $\Sigma_p^\alpha =
  \alpha^{G_p}$. Since $G_p$ is transitive on $\Sigma_p^\alpha$, it
  follows from the Orbit-Stabilizer formula that $ \# \Sigma_p^\alpha$
  divides $\# G_p$.  Hence $\# \Sigma_p^\alpha$ is $p^l$ for some $l$.
  Since $G_p\lhd G$, by Lemma~\ref{lem-orbit-normal} it follows that
  its orbit $\Sigma_p^\alpha$ is a nontrivial $G$-block. Hence $\#
  \Sigma_p^\alpha = p^l$ for some $l > 0$.  Since $p$ divides the
  cardinality of a $G$-block $\Sigma_p^\alpha$, $p$ divides $\#
  \Omega$.

  Part (2): {From} the Galois correspondence of $G$-blocks
  (Theorem~\ref{thm-blocks-galois}) we have $[\Omega :
  \Sigma_p^\alpha] = [G : G_{\Sigma_p^\alpha}]$. Notice that $p$ is
  not a factor of $[G:G_p]$ as $G_p$ is the $p$-Sylow subgroup of $G$.
  Since $G_p \lhd G_{\Sigma_p^\alpha}$ it follows that $p$ is not a
  factor of $[G : G_{\Sigma_p^\alpha}]$.  Hence $p$ is not a factor of
  $[\Omega:\Sigma^\alpha_p]$.

  Part (3): notice that $G_\Delta$ is a nilpotent group with the
  unique normal $q$-Sylow subgroup $G_q \cap G_\Delta$. Thus,
  $G_\Delta = \prod_q (G_q \cap G_\Delta)$. By
  Theorem~\ref{thm-blocks-galois}) we have
  \begin{equation}\label{eqn-index}
    \# \Delta = [G_\Delta : G_\alpha ] = \prod_q [ G_q \cap G_\Delta : 
G_q \cap G_\alpha].
  \end{equation}
  Since $G_q \cap G_\Delta$ is a $q$-group, $p$ divides $[G_q \cap
  G_\Delta : G_q \cap G_\alpha]$ if and only if $q = p$. However, in
  Equation~\ref{eqn-index}, $\# \Delta$ is a power of $p$. This forces
  $[G_q \cap G_\Delta :G_q \cap G_\alpha] = 1$ for all $q \neq p$.
  Thus $G_q \cap G_\Delta = G_q \cap G_\alpha$ for $q\neq p$.
  Therefore, $G_\Delta$ is the product group $G_p \cap G_\Delta \times
  \prod_{q \neq p} G_q \cap G_\alpha$. Since $G_{\Sigma_p^\alpha}$
  contains both $G_p$ and $G_\alpha$ we have $G_{\Sigma_p^\alpha} \geq
  G_\Delta$. Thus, $\Delta$ is a $G$-subblock of $\Sigma_p^\alpha$.
\end{proof}  

We recall a result about permutation $p$-groups (see e.g.\
Luks~\cite[Lemma 1.1]{luks82bounded}).

\begin{lemma}\label{lem-luks-pgroups}
  Let $G\leq\Sym{\Omega}$ be a transitive $p$-group and $\Delta$ be a
  maximal $G$-block.  Then $[\Omega : \Delta]=p$ and $G_\Delta =
  \Gof{\Omega/\Delta}=G^{\Delta}$ is a normal group of index $p$ in
  $G$.
\end{lemma}

The next lemma is an easy consequence of Lemma~\ref{lem-luks-pgroups}
and it states a useful property of permutation $p$-groups.

\begin{lemma}\label{lem-struct-pgroups}
  Let $H\leq\Sym{\Omega}$ be a transitive $p$-group and
  $\alpha\in\Omega$. Let $\{ \alpha \} = \Delta_0 \subset \ldots
  \subset \Delta_t = \Omega$ be any maximal chain of $H$-blocks. Then
  \begin{enumerate}
  \item $[\Delta_{i+1}: \Delta_i] = p$ for all $0 \leq i < t$.
  \item $H(\Delta_{i+1}/\Delta_i) = H_{\Delta_i}$. Hence,
    $H_{\Delta_i}\lhd H_{\Delta_{i+1}}$ and the quotient
    $H_{\Delta_{i+1}}/H_{\Delta_i}$ is cyclic of order $p$.
  \end{enumerate}
\end{lemma}

Continuing with the notation of Lemma~\ref{lem-struct-nilpotent}, we
characterize nilpotent transitive permutation groups by properties of
maximal chains of $G$-blocks between $\{ \alpha \}$ and
$\Sigma_p^\alpha$. This turns out to be crucial for our
polynomial-time nilpotence test. This characterization is probably
well-known to group theorists. However, as we haven't seen it
anywhere, we include a proof.

\begin{theorem}\label{thm-nilpotent-main-theorem}
  Let $G\leq\Sym{\Omega}$ be a transitive permutation group satisfying
  properties (1) and (2) of Lemma~\ref{lem-struct-nilpotent} (which
  are necessary conditions for nilpotence of $G$). Fix an
  $\alpha\in\Omega$. The following statements are equivalent.
  \begin{enumerate}
  \item[(1)] $G$ is nilpotent.
  \item[(2)] For each prime factor $p$ of $\# G$, every maximal chain
    of $G$-blocks $\{ \alpha \} = \Delta_0 \subset \ldots \subset
    \Delta_m = \Sigma_p^\alpha$ has the property that
    $[\Delta_{i+1}:\Delta_i] = p$, $G_{\Delta_i}$ is a normal subgroup
    of $G_{\Delta_{i+1}}$, and $p$ does not divide the order of
    $G/G^{\Delta_m}$.
  \item[(3)] For each prime $p$ dividing $\# G$, there is a maximal
    chain of $G$-blocks $\{ \alpha \} = \Delta_0 \subset \ldots
    \subset \Delta_m = \Sigma_p^\alpha$ with the property that
    $[\Delta_{i+1}:\Delta_i] = p$, $G_{\Delta_i}$ is a normal subgroup
    of $G_{\Delta_{i+1}}$, and $p$ does not divide the order of
    $G/G^{\Delta_m}$.
  \end{enumerate}
\end{theorem}

\begin{proof}
Clearly (2) implies (3). It suffices to show that (3) implies (1)
and (1) implies (2).

To see that (3) implies (1) it is enough to show that each Sylow
subgroup of $G$ is normal. To this end, let $p$ be a prime factor of
$\# G$ and let $\{ \alpha \} = \Delta_0 \subset \ldots \subset
\Delta_m = \Sigma_p^\alpha$ be a maximal chain of $G$-blocks having
the properties mentioned in (3). 

Firstly, since $\Gof{\Delta_{i+1}/\Delta_i}$ is the largest normal
subgroup of $G_{\Delta_{i+1}}$ that is contained in $G_{\Delta_i}$
(part 1 of Lemma~\ref{lem-gsupdelta}), (3) implies that $G_{\Delta_i}
= \Gof{\Delta_{i+1}/\Delta_i}$. Furthermore it follows from
Lemma~\ref{lem-gsupdelta} that there is a positive integer $l_i$ for
each $i$ such that the quotient group $G^{\Delta_{i+1}}/G^{\Delta_i}$
is embeddable in an $l_i$-fold product of copies of
$\frac{G_{\Delta_{i+1}}}{\Gof{\Delta_{i+1}/\Delta_i}} =
G_{\Delta_{i+1}}/G_{\Delta_i}$.  Since
$[G_{\Delta_{i+1}}:G_{\Delta_i}]=p$ it follows that
$G^{\Delta_{i+1}}/G^{\Delta_i}$ is a $p$-group for each $i$.  As $ \#
G^{\Delta_m}=\prod_{i=0}^{m-1} [G^{\Delta_{i+1}}:G^{\Delta_i}], $
$G^{\Delta_m}$ is also a $p$-group. Since $G^{\Delta_m}\lhd G$ and $p$
does not divide $[G:G^{\Delta_m}]$ it follows that $G^{\Delta_m}$ is a
normal $p$-Sylow subgroup of $G$. The nilpotence of $G$ follows as
this holds for all prime factors of $\# G$.

Next, we show that (1) implies (2). Suppose $G$ is nilpotent. Let $p$
be a prime factor of $\# G$ and $\alpha\in\Omega$. In the rest of this
proof let $H$ denote the $p$-Sylow subgroup $G_p$. Let $\widehat{H}$
denote the product $\prod_{q\neq p}G_q$ of all other Sylow subgroups
of $G$. Then $G=H\times\widehat{H}$. Recall that $\Sigma^\alpha_p$ is
the $H$-orbit of $\alpha$ and is therefore a block of $G$.\\ 

\noindent{\it Claim.}~~Each $\Delta\subseteq\Sigma^\alpha_p$ is a
$G$-block if and only if it is an $H$-block (in its transitive action
on $\Sigma^\alpha_p$). 

%% edited on 30/3/2006 removed some proof arguments that repeats for
%% previouly proved lemmas and inserted the appropriate refs.
%%

For the proof, note that any $G$-block
$\Delta\subseteq\Sigma^\alpha_p$ is an $H$-block. To prove the
converse consider any $H$-block
$\Sigma\subseteq\Sigma^\alpha_p$. Consider the group $G' = H_\Sigma
\times (\widehat{H} \cap G_\alpha)$. Firstly notice that the group
$G'$ is a subgroup of $G_{\Sigma_p^\alpha}$. Also since $G_\alpha$ is
nilpotent, we have $G_\alpha = H_\alpha \times (\widehat{H} \cap
G_\alpha)$. Furthermore since $\Sigma$ is a $H$-block, we have
$H_\Sigma \geq H_\alpha$. Therefore $G' \geq G_\alpha$ and by the
Galois correspondence of blocks (Theorem~\ref{thm-blocks-galois}),
$\Sigma = \alpha^{G'}$ is a $G$-block and $G_\Sigma = G'$. This proves
our claim.

The above claim implies that any maximal chain of $G$-blocks is a
maximal chain of $H$-blocks and vice-versa. Consider any maximal chain
of $G$-blocks
\[
\{\alpha\}=\Delta_0\subset\Delta_1\subset\ldots\subset
\Delta_m=\Sigma_p^\alpha.
\]
By Lemma~\ref{lem-struct-pgroups} we have $[\Delta_{i+1}:\Delta_i]=p$,
$H_{\Delta_i}\lhd H_{\Delta_{i+1}}$, and
$H_{\Delta_{i+1}}/H_{\Delta_i}$ is cyclic of order $p$.  Now,
$G_{\Delta_i}=H_{\Delta_i}\times\widehat{H}_{\Delta_i}$ and
$G_{\Delta_{i+1}}=H_{\Delta_{i+1}}\times\widehat{H}_{\Delta_{i+1}}$.
Notice that $\widehat{H}_{\Delta_i}$ is the product of $q$-Sylow
subgroups of $H_{\Delta_i}$ where $q$ varies over all prime factors of
$\# G$ different from $p$. But since $\Delta_i \subseteq
\Sigma^\alpha_p$, we have $\# \Delta_i = p^{l_i}$ for some $l_i$ and
hence from part 3 of Lemma~\ref{lem-struct-nilpotent} it follows that
$\widehat{H}_{\Delta_i}=\widehat{H}_\alpha$. Therefore
$G_{\Delta_i}\lhd G_{\Delta_{i+1}}$ and quotient group
$G_{\Delta_{i+1}}/G_{\Delta_i} \cong H_{\Delta_{i+1}}/H_{\Delta_i}$
\end{proof}

The following lemma is crucial for the nilpotence testing algorithm.
If $G$ is nilpotent then, for each prime factor $p$ of $\# G$, the
lemma implies that no matter how the maximal chain of blocks
$\Delta_i$ of Theorem~\ref{thm-nilpotent-main-theorem} is constructed,
it must terminate in $\Sigma^{\alpha}_p$.

\begin{lemma}\label{lem-nilpotent-block-increment}
  Let $G$ be a transitive nilpotent permutation group on $\Omega$. Let
  $p$ be any prime dividing $\# G$. Let $\Delta$ be any $G$-block such
  that $\# \Delta = p^l$ for some integer $l \geq 0$. Let $m$ be the
  highest power of $p$ that divides $\# \Omega$. If $l < m$ then we
  have
  \begin{enumerate}
  \item There exists a $G$-block $\Sigma$ such that $\Delta$ is a
    maximal $G$-subblock of $\Sigma$ and $[\Sigma: \Delta] = p$.
  \item For all $G$-blocks $\Sigma$ such that $\Delta$ is a maximal
    $G$-subblock of $\Sigma$ and $[\Sigma: \Delta] = p$, $G_\Delta$ is
    a normal subgroup of $G_\Sigma$.
  \end{enumerate}
\end{lemma}
\begin{proof}
  Since $\# \Delta$ is $p^l$ it follows that $\Delta$ is a
  $G$-subblock of $\Sigma_p^\alpha$
  (Lemma~\ref{lem-struct-nilpotent}). Also, $\Delta$ is a $G_p$-block
  on the transitive action of $G_p$ on $\Sigma_p^\alpha$ (as argued in
  the proof of part (3) of Theorem~\ref{thm-nilpotent-main-theorem}).
  If $l < m$ there is a $G_p$-block $\Sigma$ such that
  $\Sigma_p^\alpha \supseteq \Sigma \supset \Delta$ and $[\Sigma :
  \Delta] = p$. It follows that $\Sigma$ is a $G$-block contained in
  $\Sigma_p^\alpha$. This proves part 1.

  Let $\alpha\in\Delta$. It follows {from}
  Lemma~\ref{lem-struct-nilpotent} that for $q \neq p$ the $q$-Sylow
  subgroup of $G_\Sigma$ and $G_\Delta$ are both $G_q \cap G_\alpha$.
  Let $\widehat{G}_p$ be $\prod_{q \neq p} G_q$. The groups $G_\Sigma$
  and $G_\Delta$ are $(G_p \cap G_\Sigma) \times (\widehat{G}_p \cap
  G_\alpha)$ and $(G_p \cap G_\Delta) \times (\widehat{G}_p \cap
  G_\alpha)$ respectively. Moreover, $G_p \cap G_\Sigma$ and $G_p \cap
  G_\Delta$ are $p$-groups with index $[G_p \cap G_\Sigma : G_p \cap
  G_\Delta] = [G_\Sigma : G_\Delta] = [\Sigma : \Delta] = p$.
  Therefore, $G_p \cap G_\Delta$ is normal in $G_p \cap G_\Sigma$.
  Thus, $G_\Delta = (G_p \cap G_\Delta)\times (\widehat{G}_p \cap
  G_\alpha)$ is normal in $G_\Sigma = (G_p \cap G_\Sigma) \times
  (\widehat{G}_p \cap G_\alpha)$ and $\frac{G_\Sigma}{G_\Delta} =
  \frac{G_p \cap G_\Sigma}{G_p \cap G_\Delta}$ is isomorphic to
  $\mathbb{Z}_p$.
  \end{proof}

\subsection{\bf The nilpotence test}

Given $f(X) \in \Q[X]$ our goal is to test if $\Gal{f}$ is nilpotent.
We can assume that $f(X)$ is irreducible. For, otherwise we can
compute the irreducible factors of $f(X)$ over $\Q$ using the LLL
algorithm, and perform the nilpotence test on each distinct
irreducible factor. This suffices because nilpotent groups are closed
under products and subgroups. Let $G$ be $\Gal{f}$. We consider $G$ as
a subgroup of $\Sym{\Omega}$, where $\Omega$ is the set of roots of
$f(X)$. Since $f$ is irreducible, $G$ is transitive on $\Omega$.

For any $G$-block $\Delta$, let $\Q_\Delta$ be the fixed field of the
splitting field $\Q_f$ under the automorphisms of $G_\Delta$. Let
$\Delta$ be a $G$-block containing $\alpha$. Since $G_\Delta \geq
G_\alpha$, $\Q_\Delta$ is a subfield of $\Q_{\{\alpha\}} =
\Q(\alpha)$.

We describe the main idea. By
Theorem~\ref{thm-nilpotent-main-theorem}, $G$ is nilpotent if and only
if for all primes $p$ that divide the order of $G$, there is a maximal
chain of $G$-blocks $\{\alpha \} = \Delta_0\subset \ldots \subset
\Delta_m$ satisfying conditions of part (3) of
Theorem~\ref{thm-nilpotent-main-theorem}. We show these conditions can
be verified in polynomial time once the tower of fields $\Q(\alpha) =
\Q_{\Delta_0} \supset \ldots \supset \Q_{\Delta_m}$ are known. Thus,
for testing nilpotence of $G$ we will first need a polynomial-time
algorithm that computes $\Q_{\Delta_i}$. We describe this in the
following theorem.

%(We note, however, that there is no known polynomial-time
% algorithm for computing the groups $G_{\Delta_i}$.)

\begin{theorem}[Proof in Appendix]\label{thm-enlarge-block}
  Let $f(X)\in\Q[X]$ be irreducible, $G=\Gal{f}$ be its Galois group
  and $\Omega$ be the set of roots of $f$. Let $\Delta\subseteq\Omega$
  be any $G$-block and $\alpha\in \Delta$. There is an algorithm that
  given a primitive polynomial $\mu_\Delta(X) \in \Q[X]$ of
  $\Q_\Delta$, runs in time polynomial in $\size{f}$ and
  $\size{\mu_\Delta}$ and computes a primitive polynomial
  $\mu_\Sigma(X) \in \Q[X]$ of $\Q_\Sigma$ for all $G$-blocks $\Sigma$
  such that $\Delta$ is a maximal block of $\Sigma$.  Moreover
  $\size{\mu_\Sigma}$ is at most a polynomial in $\size{f}$ and is
  independent of $\size{\mu_\Delta}$.
\end{theorem}

Algorithm~\ref{algo-nilpotent} describes the nilpotence test.

\begin{algorithm}
  \SetKw{Print}{print}
  \caption{Nilpotence test}\label{algo-nilpotent}
  \KwIn{A polynomial $f(X) \in \Q[X]$ of degree $n$} \KwOut{``Accept''
    if $\Gal{f}$ is nilpotent;``Reject'' otherwise} 
 Verify that $f(X)$ is solvable\;

  \lnl{step-compute-ps} Compute the set $P$ of all the prime factors
  of $\# \Gal{f}$\;

  Let $G\leq\Sym{\Omega}$ denote the Galois group of $f$, where
  $\Omega$ is the set of roots of $f$.
  
  \lnl{step-for-loop} 
  \For{\KwSty{every} $p \in P$} {

    \If{$p$ does not divide $n$}{ \Print{Reject}}
    
    Let $m$ be the highest power of $p$ dividing $n$.
  
    \lnl{step-tower} Attempt to compute the tower
    $\Q_{\Delta_m} \subset \ldots \subset
    \Q_{\Delta_0}$ for a maximal chain of $G$-blocks
    $\{\alpha\} = \Delta_0 \subset \ldots \subset\Delta_m$ such that
    $[\Q_{\Delta_{i+1}}:\Q_{\Delta_i}] = p$.

   \lnl{step-if-normal}
   \If{Step~\ref{step-tower} fails \KwSty{or}
      $\Q_{\Delta_{i+1}}$ is not normal over
     $\Q_{\Delta_i}$}{ \Print{Reject}} 

    Let $\mu_{\Delta_m}(X)$ be the primitive polynomial for
    $\Q_{\Delta_m}$

    \lnl{step-check-psylow} 
    \If{ $p$ divides $\#\Gal{\mu_{\Delta_m}}$}
    { 
        \Print{Reject}
    }
  
  }
  
  \Print{Accept}

\end{algorithm}

%\end{proof}

We prove that Algorithm~\ref{algo-nilpotent} runs in polynomial time.
For the steps~\ref{step-compute-ps} and \ref{step-check-psylow} note
that for polynomials $f$ with solvable Galois groups, as a byproduct
of the Landau-Miller test~\cite{landau85solvability}, the prime
factors of $\# \Gal{f}$ can be found in polynomial time (see also
Theorem~\ref{thm-gammad-primes}). We explain how step~\ref{step-tower}
can be done in polynomial time using Theorem~\ref{thm-enlarge-block}.
We construct $\Q_{\Delta_i}$ inductively starting with $\Q_{\Delta_0}
= \Q(\alpha)$. Assume we have computed $\Q_{\Delta_i}$.  Using
Theorem~\ref{thm-enlarge-block} we compute $\Q_\Sigma$ for each
$G$-block $\Sigma$ containing $\Delta_i$ as a maximal $G$-subblock.
Among them choose a $\Q_\Sigma$ for which $[\Sigma : \Delta_i] = p$
and let $\Q_{\Delta_{i+1}}$ be $\Q_\Sigma$. The inductive construction
of $\Q_{\Delta_{i+1}}$ from $\Q_{\Delta_i}$ can be done in time
bounded by a polynomial in $\size{f}$. Putting it together we have the
following proposition.

\begin{proposition}
  Algorithm~\ref{algo-nilpotent} runs in time polynomial in
  $\size{f}$.
\end{proposition}

We now argue its correctness. Part (1) of
Theorem~\ref{thm-nilpotent-main-theorem} implies that if $G$ is
nilpotent then Algorithm~\ref{algo-nilpotent} accepts. Conversely,
suppose the algorithm accepts. Then for each prime $p$ dividing $\# G$
we have a maximal chain of $G$-blocks $\{ \alpha \} = \Delta_0 \subset
\ldots \subset \Delta_m$ such that $\Q_{\Delta_i}/\Q_{\Delta_{i+1}}$
are normal extensions for each $0 \leq i < m$ (this we verify in
step~\ref{step-if-normal} of Algorithm~\ref{algo-nilpotent}). Recall
that $\Q_{\Delta_i}$ is the fixed field of $\Q_f$ w.r.t.\
$G_{\Delta_i}$. Hence by checking $\Q_{\Delta_i}/\Q_{\Delta_{i+1}}$ is
a normal extension we have verified that $G_{\Delta_i}\lhd
G_{\Delta_{i+1}}$. Also, the splitting field of the primitive
polynomial $\mu_{\Delta_m}(X)$ is the normal closure of
$\Q_{\Delta_m}$ over $\Q$. It follows from Lemma~\ref{lem-gsupdelta}
and Theorem~\ref{thm-funda-galois} that $\Gal{\mu_{\Delta_m}}$ is
$G^{\Delta_m}$. Hence, by checking $p$ does not divide $\#
\Gal{\mu_\Delta}$ we have verified that $p$ does not divide $\#
G/G^{\Delta_m}$. Thus, we have verified that the maximal chain of
$G$-blocks $\{ \alpha \} = \Delta_0 \subset \ldots \subset \Delta_m$
satisfies the conditions of Part(3) of
Theorem~\ref{thm-nilpotent-main-theorem} implying that $G$ is
nilpotent. Putting it all together we have the following theorem.

\begin{theorem}
  There is a polynomial-time algorithm that takes $f\in\Q[X]$ as input
  and tests if $\Gal{f}$ is nilpotent.
\end{theorem}

\section{Generalizing the Landau-Miller solvability test}

In this section we show that the Landau-Miller solvability test can be
adapted to test if the Galois group of $f(X) \in \Q[X]$ is in
$\Gamma_d$ for constant $d$. Note that for $d < 5$, $\Gamma_d$ is the
class of solvable groups and hence our result is a generalization of
the result of Landau-Miller~\cite{landau85solvability}. We first
recall a well-known bound on the size of primitive permutation groups
in $\Gamma_d$.
\begin{theorem}[\cite{babai82primitive}]\label{thm-babai-cameron-palfy}
  Let $G \leq S_n$ be a primitive permutation group in $\Gamma_d$ for
  a constant $d$. Then $\# G \leq n^{O(d)}$.
\end{theorem}

\begin{theorem}\label{thm-gammad-test}
  For constant $d>0$, there is an algorithm that takes as input $f(X)
  \in \Q[X]$ and in time polynomial in $\size{f}$ and $n^{O(d)}$
  decides whether $\Gal{f}$ is in $\Gamma_d$.
\end{theorem}
\begin{proof}
  We sketch the proof. Assume without loss of generality that $f(X)$
  is irreducible.  Let $G=\Gal{f}$ as a subgroup of $\Sym{\Omega}$,
  where $\Omega$ is the set of roots of $f$. Let $\{ \alpha \} =
  \Delta_0 \subset \ldots \subset \Delta_t = \Omega$ be any maximal
  chain of $G$-blocks. The series $ \{ 1 \} = G^{\Delta_0} \lhd \ldots
  \lhd G^{\Delta_t} = G$ gives a normal series for $G$. By closure
  properties of $\Gamma_d$, $G\in \Gamma_d$ iff
  $\frac{G^{\Delta_{i+1}}}{G^{\Delta_i}}\in\Gamma_d$ for each $i$. If
  $G$ is in $\Gamma_d$ so are $G_{\Delta_{i+1}}$ and
  $\Gof{\Delta_{i+1}/\Delta_i}$ and hence their quotient
  $\frac{G_{\Delta_{i+1}}}{\Gof{\Delta_{i+1}/\Delta_i}}$. On the other
  hand since $\frac{G^{\Delta_{i+1}}}{G^{\Delta_i}}$ is isomorphic to
  a subgroup of
  $\left(\frac{G_{\Delta_{i+1}}}{\Gof{\Delta_{i+1}/\Delta_i}}\right)^l$
  for some $l$ (Lemma~\ref{lem-gsupdelta}),
  $\frac{G^{\Delta_{i+1}}}{G^{\Delta_i}}\in\Gamma_d$ if
  $\frac{G_{\Delta_{i+1}}}{\Gof{\Delta_{i+1}/\Delta_i}}\in\Gamma_d$.
  Hence $G\in\Gamma_d$ iff
  $\frac{G_{\Delta_{i+1}}}{\Gof{\Delta_{i+1}/\Delta_i}}$ is in
  $\Gamma_d$ for each $i$ . We give a polynomial-time algorithm to
  verify the above fact for some maximal chain of $G$-blocks $\{
  \alpha \} = \Delta_0 \subset \ldots \subset \Delta_t = \Omega$.

  First, by Theorem~\ref{thm-enlarge-block}we compute $K_i =
  \Q_{\Delta_i}$ for a maximal chain of $G$-blocks $\{ \alpha \} =
  \Delta_0 \subset \ldots \subset \Delta_t = \Omega$. Let $L_i$ be the
  fixed field of $\Q_f$ with respect to the automorphisms of
  $\Gof{\Delta_{i+1}/\Delta_i}$ then $L_{i+1}$ is the normal closure
  of $K_i$ over $K_{i+1}$. This follows because
  $\Gof{\Delta_{i+1}/\Delta_i}$ is the largest proper normal subgroup
  of $G_{\Delta_{i+1}}=\Gal{\Q_f/\Q_{\Delta_{i+1}}}$. Hence
  $\Gal{L_{i+1}/K_{i+1}}$ is
  $\frac{G_{\Delta_{i+1}}}{\Gof{\Delta_{i+1}/\Delta_i}}$, and it
  suffices to check that each $\Gal{L_i/K_i}$ is in $\Gamma_d$.

  The group $\frac{G_{\Delta_{i+1}}}{\Gof{\Delta_{i+1}/\Delta_i}}$
  acts faithfully and primitively on $\Omega^\prime =
  \Blocks{\Delta_{i+1}/\Delta_i}$, by Lemma~\ref{lem-gsupdelta} and
  since $\Delta_i$ is a maximal subblock of $\Delta_{i+1}$. If
  $G\in\Gamma_d$ then $[L_{i+1}:K_{i+1}] = \# \Gal{L_{i+1}/K_{i+1}}
  \leq n^{O(d)}$ and degrees $[L_i: \Q]$ are all less than $n^{O(d)}$.
  We can use Theorem~\ref{thm-landau-galois-algo} to compute
  $\Gal{L_i/K_i}$ as a multiplication table in time polynomial in
  $\size{f}$ and $n^d$ for each $i$. We then verify that
  $\Gal{L_i/K_i}\in\Gamma_d$ by computing a composition series for it
  and checking that each composition factor is in $\Gamma_d$. At any
  stage in the computation of $\Gal{L_i/K_i}$ if the sizes of the
  fields becomes too large, i.e.  larger than the bound of
  Theorem~\ref{thm-babai-cameron-palfy} we abort the computation and
  decide that $\Gal{f}$ is not in $\Gamma_d$. Clearly, these steps can
  be done in polynomial time.
\end{proof}

It follows from the proof of Theorem~\ref{thm-gammad-test} that a
prime $p$ divides $\# \Gal{f}$ if and only if it divides $[L_i: K_i]$
for some $1 \leq i \leq t$. Hence we have the following theorem.

\begin{theorem}\label{thm-gammad-primes}
  Given $f(X) \in \Q[X]$ with Galois group in $\Gamma_d$ there is an
  algorithm running in time polynomial in $\size{f}$ and $n^d$ that
  computes all the prime factors of $\# \Gal{f}$.
\end{theorem}

%\section{Concluding Remarks}
%We have described a polynomial-time algorithms for testing if the
%Galois group of a polynomial $f(x)\in\Q[X]$ is nilpotent or in the
%class $\Gamma_d$. An interesting problem along the same lines is
%whether we can efficiently test if the Galois group is supersolvable.
%Supersolvable groups are a proper subclass of solvable groups and
%contain nilpotent groups. Using the same ideas, we can do
%polynomial-time nilpotence testing of the Galois group of $f(X)\in
%K[X]$, where $K$ is any number field presented by giving a primitive
%polynomial $\mu(X)$ of $K$ over $\Q$.

\appendix 
\newpage

\section{Appendix: Proof of Theorem~\ref{thm-enlarge-block}}

Let $f(X)$ be an irreducible polynomial and let $G$ be its Galois
group thought of as permutation group over $\Omega$, the set of roots
of $f(X)$.  For a $G$-block $\Delta$ let $T_\Delta(X)$ be the
polynomial defined by
\[
T_\Delta(X) = \prod_{\eta \in \Delta} (X - \eta).
\]
Note that if $\Delta \ni \alpha$ then $T_\Delta(X) \in
\Q(\alpha)[X]$ and the field
$\Q(\delta_0,\ldots,\delta_r) = \Q_{\Delta}$. Hence
computing $\Q_\Delta$ reduces to computing $T_\Delta$.

\begin{lemma}\label{lem-factor-blocks}
  Let $\Delta$ be a $G$-block containing $\alpha$. The irreducible
  factor of $f$ over $\Q_\Delta$ which has $\alpha$ as root is
  $T_\Delta$. Let $\Sigma$ be any $G$-block such that $\Sigma
  \supseteq \Delta$. If $g$ is an irreducible factor of $f$ over
  $\Q_\Delta$ then $\Sigma$ contains a root of $g$ if and only
  if it contains all the roots of $g$.
\end{lemma}
\begin{proof}
  Let $g$ be an irreducible factor of $f(X)$ over $\Q_\Delta$.
  The roots of $g$ forms a $G_\Delta$-orbit of $\Omega$. Conversely for
  any $G_\Delta$-orbit $\Omega^\prime$ the polynomial $\prod (X -
  \eta)$, $\eta$ varies over $\Omega^\prime$, is an irreducible factor
  of $f(X)$ over $\Q_\Delta$. Hence the irreducible factor
  that contains $\alpha$ as root is $T_\Delta$.

  For a $G$-block $\Sigma$ containing $\Delta$ we have $G_\Sigma \geq
  G_\Delta$. Hence for any irreducible factor $g(X)$ of $f(X)$ over
  $\Q_\Delta$ two roots $\eta_1$ and $\eta_2$ of $g(X)$ are in
  the same $G_\Sigma$ orbit.  Hence $\eta_1 \in \Sigma$ if and only if
  $\eta_2 \in G_\Sigma$ as $\Sigma$ is the $G_\Sigma$-orbit
  $\alpha^{G_\Sigma}$.
\end{proof}

Let $\Delta$ be a $G$-block containing $\alpha$ and assume that we
know $\Q_\Delta$. Assume that $f$ factors as $g_0\ldots g_r$
over $\Q_\Delta$. One of these factors say $g_0$ is
$T_\Delta$. Consider any $G$-block $\Sigma$ such that $\Delta$ is a
maximal $G$-subblock of $\Sigma$. There is a factor $g_i$ such that
$\Sigma$ contains a root (hence all the roots by
Lemma~\ref{lem-factor-blocks}) $g_i$.  Let $\Sigma_i$ be the smallest
$G$-block containing $\Delta$ and all the roots of $g_i$.  We give a
polynomial time algorithm to compute $T_{\Sigma_i}$.
Theorem~\ref{thm-enlarge-block} then follows from this algorithm.

\begin{lemma}\label{lem-enlarge-block}
  Let $\Delta$ be a $G$-block containing $\alpha$. Given the field
  $\Q_\Delta$ as a subfield of $\Q(\alpha)$ and an
  irreducible factor $g$ of $f$ over $\Q_\Delta$ we can
  compute in polynomial time $T_\Sigma$ as a polynomial in
  $\Q(\alpha)[Y]$, where $\Sigma$ is the smallest $G$-block
  containing $\Delta$ and the roots of $g$. 
\end{lemma}
\begin{proof}
  We are given $\Q_\Delta$ as a subfield of
  $\Q(\alpha)$. We compute a primitive element $\eta$ of
  $\Q_\Delta$ as a polynomial in $\alpha$.  The coefficients
  of factors of $f$ over $\Q_\Delta$ are polynomials in
  $\eta$. Let the factorisation of $f$ over $\Q_\Delta$ be
  $f=g_0 \ldots g_r$, where $g_0 = T_\Delta$ and $g=g_1$.  Denote the
  set of roots of $g_i$ by $\Phi_i$, for each $i$. Then $\Phi_i$'s are
  the orbits of $G_\Delta$ and by Lemma~\ref{lem-factor-blocks}, the
  polynomial $T_\Sigma$ is precisely the product of $g_i$ such that
  $\Phi_i\subseteq \Sigma$.
  
  Let $\beta$ denote a root of $g(X)$, and
  $\sigma \in \Gal{\Q_f/\Q}$ be an automorphism such
  that $\sigma$ maps $\alpha$ to $\beta$. Notice that $\sigma$ is an
  isomorphism between the fields $\Q(\alpha)$ and
  $\Q(\beta)$.  Let $\Sigma$ be the smallest $G$-block
  containing $\Delta$ and $\Phi_1$. {From}
  Theorem~\ref{thm-blocks-galois} and the Galois correspondence of
  blocks (Theorem~\ref{thm-blocks-galois}) we know that $G_\Sigma$ is
  generated by $G_\Delta\cup\{\sigma\}$.
 % By Lemma~\ref{lem-factor-blocks}, $\Sigma$ is the union of $\Delta$
 % and some of the orbits $\Phi_i$.
  We can find these orbits by the following transitive closure kind of
  procedure (its correctness follows directly from
  Lemma~\ref{lem-factor-blocks}).

  \begin{algorithm}
    \caption{Computing $\Sigma$}%
    \label{alg-compute-sigma}
    Let $S:= \{\Delta,\Phi_1\}$ 

    \While{new orbits get added to $S$} {%
      Compute $T:=\{\Phi^\sigma\mid \Phi\in S\}$
      
      \lIf{$\Phi_j\cap \Phi^\sigma\neq\emptyset$ for some
	$\Phi^\sigma\in T$} % 
	  {include $\Phi_j$ in $S$}% 
    }

    Output $\bigcup\{\Phi\mid \Phi\in S\}$
    
  \end{algorithm}

  Our goal is to get a polynomial-time algorithm for computing
  $T_\Sigma$ from the above procedure that defines $\Sigma$.  First,
  we compute the extension field $\Q(\alpha,\beta)=\Q(\gamma)$: we do
  this by first factoring $f$ over $\Q(\alpha)$. Let $h$ be an
  irreducible factor of $g$ over $\Q(\alpha)$. Then
  $\Q(\alpha,\beta)=\Q (\alpha)[X]/h(X)$. As
  $[\Q(\alpha,\beta):\Q]\leq n^2$, we can compute a primitive element
  $\gamma$ in polynomial time.  Furthermore, in polynomial time we
  will find polynomials $r_1$ and $r_2$ such that $\alpha=r_1(\gamma)$
  and $\beta=r_2(\gamma)$.

  Let $\sigma$ map the polynomials $g_0,\ldots,g_r$ in
  $\Q(\alpha)[X]$ to the polynomials
  $g_0^\sigma,\ldots,g_r^\sigma$ in $K(\beta)[X]$, obtained by
  symbolically replacing $\alpha$ by $\beta$ in each coefficient of
  the polynomials $\{ g_i : 0 \leq i \leq r\}$'s.

  In Algorithm~\ref{alg-compute-sigma}, testing if
  $\Phi_j\cap\Phi_i^\sigma\neq\emptyset$ amounts to finding if
  $gcd(g_j,g_i^\sigma)$ is nontrivial. To make this gcd computation
  possible, we must express $g_j$ and $g_i$ over $K(\gamma)$, which we
  do by replacing $\alpha$ by $r_1(\gamma)$ and $\beta$ by
  $r_2(\gamma)$.  We can now give the algorithm for computing
  $T_\Sigma$.
  \begin{algorithm}
    \caption{Computing $T_\Sigma$}%
    \label{alg-enlarge-block}
    Let $S:= \{T_\Delta,g\}$% 
    \While{new factors get included in $S$}%
	  { %

	    Compute $S' := \{
	    g_i^\sigma\mid g_i\in S\}\cup\{T_\Delta^\sigma\} $\
	    
	    \For{each factor $g_j$}
		{%
		  
		  \lIf{$gcd(g_j,h')$ is nontrivial for
		    some $h'\in S'$}% 
		      {include $g_j$ in $S$}%
		}%
		  
		  /* Notice that the gcd computation is done by */%
		  /* expressing $g_j$ and $h'$ over $K(\gamma)$ */\\ 
		}%

		Output $T_\Sigma:=T_\Delta\cdot\prod_{g_i \in S}
    g_i$
  \end{algorithm}
  
  It is clear that Algorithm~\ref{alg-enlarge-block} is
  polynomial-time bounded. The preceding discussion and the procedure
  for defining $\Sigma$ imply that the algorithm correctly computes
  $T_\Sigma$.
\end{proof}
\end{document}